\documentclass[preprint,aps,twocolumn,10pt]{revtex4}
\usepackage{amsmath,amssymb,amsthm,graphicx,latexsym}
\newcommand{\doublespacing}{\let\CS=\@currsize\renewcommand{\baselinesstrech}
{2.0}\tiny\CS}

\newcommand{\bd}{\begin{document}}
\newcommand{\ed}{\end{document}}
\newcommand{\bc}{\begin{center}}
\newcommand{\ec}{\end{center}}
\newcommand{\vs}{\vspace}

\begin{document}

\title{\Large \bf Scattering states of a particle, with position-dependent
mass, in a double heterojunction}

\vs{.3cm}

\author{{\large Anjana Sinha}$^*$ \\
Department of Applied Mathematics, Calcutta University, 92 A.P.C.
Road, Kolkata - 700 009, INDIA }

\vs{1cm}


\begin{abstract}

\noindent In this work we obtain the exact analytical scattering
solutions of a particle (electron or hole) in a semiconductor
double heterojunction --- potential well / barrier --- where the
effective mass of the particle varies with position inside the
heterojunctions. It is observed that the spatial dependence of
mass within the well / barrier introduces a nonlinear component in
the plane wave solutions of the continuum states. Additionally,
the transmission coefficient is found to increase with increasing
energy, finally approaching unity, whereas the reflection
coefficient follows the reverse trend and goes to zero.

\vs{.3cm}


\vs{.3cm}

\noindent $^*$ e-mail : anjana23@rediffmail.com

\end{abstract}


\maketitle

\newpage

\section{Introduction}

Recent developments in the nanofabrication of semiconductor
devices have given a thrust to the study of quantum mechanical
systems with position dependent effective mass. The spatial
dependence on the effective mass of the particle arises due to its
interaction with an ensemble of particles within the device, as
the particle propagates from left to right. This so-called
position-dependent effective mass (pdem) formalism becomes an
essential ingredient in describing the electronic and transport
properties of quantum wells and quantum dots, impurities in
crystals, He-clusters, quantum liquids, semiconductor
heterostructures, etc.
\cite{pdm1,pdm2,pdm3,pdm4,pdm5,harrison,bastard,levy-leblond}.
When two such materials (having pdem) with different bandgaps are
placed adjacent to each other to form a heterojunction, the
effective mass approximation is valid within each material. If,
for example, a thin layer of a narrower (wider) bandgap material
is sandwiched between two layers of a wider (narrower) bandgap
material, they form a double heterojunction. If the intermediate
layer is sufficiently thin for quantum properties to be exhibited,
then the alignment is called a single quantum well (barrier). A
typical quantum well structure may be composed of a semiconductor
thin film, embedded between two semi-infinite semiconductor
materials, say $GaAs/Al_{x}Ga_{1-x}As$, where $x$ denotes the mole
fraction. As the mole fraction varies along the $z$-axis, so does
the effective mass of the charge carrier (electron or hole). To
have a complete understanding of such a quantum system, quantum
mechanics requires knowledge of both bound and scattering states.

Various attempts have been made over the years to study the
scattering states in such position-dependent effective mass
systems
\cite{pra75,pra83,pra59,scatt-pdm,koc-koca,paranjpe,jha,jpa43}. In
two of the recent works \cite{prb83,prb82}, the authors studied
the effect of hard wall confinement and lateral dimensions on low
temperature transport properties of long diffuse channels in
$InSb/In_{1-x}Al_x Sb$ heterostructures \cite{prb83}, and resonant
inelastic x-ray scattering in $LaAlO_3/SrTiO_3$ heterostructures
\cite{prb82}. In another work, ballistic carrier emission was
studied with $GaAs/Al_{x}Ga_{1-x}As$ as a model system
\cite{prb81}. In one of the relatively earlier works, the authors
considered the simple model of a square potential well, with the
effective mass varying with position inside the well
\cite{koc-koca}. However, as the additional term introduced by the
changing mass is small compared with the original potential $V_0$
and does not change the shape of the potential significantly, this
term was neglected while finding the solutions.

In the present work our aim is to obtain the exact analytical
solutions for the scattering states of a particle inside a single
quantum well/barrier. It may be mentioned that exact analytical
solutions play an important role in conceptual understanding of
physics. They provide a valuable platform for checking and
improving approximate models and numerical results. Herein lies
the motivation for obtaining exact analytical solutions of wave
equations with pdem, especially because of the wide range of
applications of these solutions in various  areas of material
science and condensed matter \cite{alhaidari}. In the double
heterojunction considered here, we assume the intermediate layer
to be a potential well / barrier of the form
\begin{equation}\label{pot-form}
        V = \left\{
    \begin{array}{lcl}
        & & \displaystyle V(z)
        \qquad \ \ , \ \ \ a_1 < z < a_2 \\
        & & \displaystyle V_{01} = V(a_1) \  , \ \ \ - \infty < z < a_1 \\
        & & \displaystyle V_{02} = V(a_2) \  , \ \ \ a_2 < z < \infty
    \end{array}
        \right.
\end{equation}
where $a_1$ and $a_2$ represent the heterojunctions. The mass of
the charge carrier is assumed to be spatially varying inside the
potential well / barrier $a_1 < z < a_2$, but constant outside,
viz,
\begin{equation}\label{mass-form}
        m = \left\{
    \begin{array}{lcl}
        & & \displaystyle m(z)
        \qquad \ \ , \ \ \ a_1 < z < a_2 \\
        & & \displaystyle m_1 = m(a_1)  \ , \ \ \ - \infty < z < a_1 \\
        & & \displaystyle m_2 = m(a_2)  \ , \ \ \ a_2 < z < \infty
    \end{array}
        \right.
\end{equation}
Thus both the potential $V(z)$ and mass $m(z)$ are real functions
of the configuration space coordinates, and are taken to be
continuous throughout the semiconductor device. The work done in
refs. \cite{jha, alhaidari} deserve special mention here. In ref.
\cite{jha} approximate analytical solutions were derived for any
arbitrary potential and arbitrary mass function, whereas in ref.
\cite{alhaidari}, some special forms of the mass function were
considered for oscillator, Coulomb and Morse potentials to produce
exact analytical results. However, our study differs significantly
from both these works --- the mass functions considered in our
case, as well as the approach used, are different from refs.
\cite{jha} and \cite{alhaidari}; the present article not only
gives exact analytic results but also plots the transmission and
reflection coefficients. More importantly, unlike
\cite{jha,alhaidari}, this work deals with a double
heterojunction.

\vs{.2cm}

The article is organized as follows : For the sake of
completeness, the position-dependent-mass Schr\"{o}dinger equation
is introduced in Section II, and the method of obtaining the
solutions is discussed. A couple of explicit models are studied in
Section III. To give a better insight into the physical nature of
the problem, the potential and mass functions are plotted as a
function of $z$ (in Fig. 1 and Fig. 3) along with the scattering
solutions (in Fig. 2 and Fig. 4). The transmission and reflection
coefficients are also calculated and the same are plotted as a
function of the energy of the particle, in Fig. 5 and Fig. 6
respectively. Section IV is kept for Conclusions and Discussions.

\section{Theory}

\noindent We start with the basic one dimensional time independent
Schr\"{o}dinger equation associated with a particle endowed with
pdem in the intermediate region within the heterojunctions :
\begin{equation}\label{H-em}
    \displaystyle H_{EM} (z) \psi (z)  \equiv  \left[ T_{EM} (z) + V(z) \right]
    \psi (z) = E \psi (z)
\end{equation}
where the kinetic energy term $T_{EM}$ is given by
\cite{pdm4,harrison}
\begin{equation}\label{T-em}
    \begin{array}{lcl}
    T_{EM} &=& \displaystyle \frac{1}{4} \left( m^{\alpha} p
    m^{\beta} p m ^{\gamma} + m^{\gamma} p
    m^{\beta} p m ^{\alpha} \right) \\ \\
    &=& \displaystyle \frac{1}{2} p \left( \frac{1}{m}
    \right) p
    \end{array}
\end{equation}
with $ p = \displaystyle - i \hbar \frac{d}{dz} $ being the
momentum operator, and the ambiguity parameters $\alpha \ , \
\beta \ , \ \gamma $ obeying the von Roos constraint \cite{pdm4}
    \begin{equation}\label{abg}
        \alpha + \beta + \gamma = -1
    \end{equation}
There is neither a unique nor a universal choice for the ambiguity
parameters, and several suggestions exist in literature --- e.g. $
\alpha = \gamma = 0 \ , \ \beta = -1$ \cite{benDaniel-Duke}, $
\alpha = \gamma = - 1/2 \ , \ \beta = 0$ \cite{z-k}, $ \alpha =
\gamma = - 1/4 \ , \ \beta = - 1/2$ \cite{m-m}, $ \beta = \gamma =
- 1/2 \ , \ \alpha = 0$ \cite{l-k}, etc. However, for continuity
conditions at the abrupt interfaces, one should consider $ \alpha
= \gamma$, or else one gets the unphysical result of the wave
function vanishing at the heterojunctions; additionally the ground
state energy also diverges \cite{marrow, thomsen}. We shall
restrict ourselves to the BenDaniel-Duke choice for the ambiguity
parameters, viz., $ \alpha = \gamma = 0 \ , \ \beta = -1 $.
Incidentally, this particular choice consistently produces the
best fit to experimental results \cite{proceed}. Furthermore, we
shall work in units $\hbar = c = 1$, and use prime to denote
differentiation w.r.t. $z$. Thus, inside the potential
well/barrier $a_1 < z < a_2$, the Hamiltonian for the particle
with pdem reduces to \cite{plastino}
\begin{equation}\label{h-in}
        H = \displaystyle - \frac{1}{2m(z)} \frac{d^2}{dz^2} - \left(
        \frac{1}{2m(z)} \right) ^{\prime} \frac{d}{dz} + V(z)
\end{equation}
whereas, outside the well/barrier, $ z < a_1 $ and $ z > a_2 $,
the particle obeys the conventional Schr\"{o}dinger equation :
\begin{equation}\label{sch-out}
    \displaystyle \left\{ - \frac{1}{2m_{1,2}} \frac{d^2}{dz^2} +
    V_{01,02}
    \right\} \psi (z)  = E \psi (z)
\end{equation}
having plane wave solutions. In case we consider a wave incident
from left, the solutions in the two regions are
\begin{equation}\label{psi-out}
\begin{array}{lcl}
    \psi _L (z)  &=& \displaystyle e^{i k_1 z} + R e^{-ik_1 z} \ , \ - \infty <
    z < a_1 \\ \\
    \psi _R (z)  &=& T e^{i k_2 z} \ , \ \qquad \qquad a_2 <
    z < \infty \\
\end{array}
\end{equation}
where $R$ and $T$ denote the reflection and transmission
amplitudes, and
\begin{equation}\label{k}
    k_{1,2} = \displaystyle  \sqrt{ 2 m_{1,2} \left( E - V_{01,02} \right) }
\end{equation}

\vs{.1cm}

\noindent To find the solution in the region $a_1 < z < a_2$, we
make use of the following transformations \cite{br-pr}
    \begin{equation}\label{psi-z}
        \psi _{in} = \displaystyle \left\{ 2 m(z) \right\} ^{1/4} \phi
         \ , \ \rho = \displaystyle \int \sqrt{2 m(z)} dz
    \end{equation}
which reduce the Schr\"{o}dinger equation for position-dependent
mass, to one for constant mass, viz.,
\begin{equation}\label{schro-const-m}
    \displaystyle -  \frac{d^2 \phi }{d \rho ^2} + \left\{
    \widetilde{V} (\rho) - E \right\} \phi  = 0
\end{equation}
with
\begin{equation}\label{v-tilde}
    \widetilde{V} (\rho) = \displaystyle V(z) + \frac{7}{32} \frac{m^{\prime
    \ 2}}{m^3} - \frac{m^{\prime \prime}}{8 m^2}
\end{equation}
We are interested in studying the scattering states of a particle
in a double heterojunction, formed by dissimilar materials, where
the mass of the particle varies with position inside the well /
barrier. Such heterojunctions can be described by a material
potential which derives from the difference in bandgaps
\cite{harrison}. Crystal potential of multiple heterojunction can
also be described in this manner. For this purpose, we look for
some definite practical forms of $V(z)$ and $m(z)$ which will give
exact analytical solutions of (\ref{schro-const-m}). We illustrate
this with the help of a couple of explicit models in the next
section.

\section{Explicit models}

\subsection{Case 1 : Potential well with position dependent mass}

\noindent The one dimensional finite square well is one of the
simplest confinement potentials. In the first example, we consider
the region between the abrupt heterojunctions ($-a_0 < z < a_0$)
to be a symmetric potential well (more realistic than the square
well) with the following ansatz ($\mu$ being some constant)
\begin{equation}\label{pot-1}
        V(z) = \left\{
    \begin{array}{lcl}
        & & \displaystyle - \ \frac{\mu ^2 }{1 + z^2}
        \ \  , \ -a_0 < z < a_0 \ ,  \\ \\
        & & \displaystyle - \frac{\mu ^2 }{1 + a_0^2} \ = \
        V_0 \ , \\
        & & \ \ - \infty < z < -a_0 \ ,  \ a_0 < z < \infty
    \end{array}
    \right.
\end{equation}
This particular model resembles the profile of a diffused quantum
well, with the advantage of exact analytical solutions. In case
the real situation is slightly different from this model, one can
apply approximation methods like perturbation theory, etc.,
to obtain the solutions. \\
Let the mass of the particle be
\begin{equation}\label{mass-1}
        m(z) = \left\{
    \begin{array}{lcl}
        & & \displaystyle \frac{\beta ^2}{2 \left(1 + z^2 \right)}
        \ \ \ , \ -a_0 < z  < a_0 \ , \\ \\
        & & \displaystyle \frac{\beta ^2}{2 \left(1 + a_0^2 \right)} = m_0 \
        , \\
        & & \ \ - \infty < z < -a_0 \ , \ a_0 < z < \infty \\
    \end{array}
        \right.
\end{equation}
where $\beta$ is some constant parameter. For the spatial mass
dependence given by eq. (\ref{mass-1}), eq. (\ref{psi-z})
transforms the coordinate $z$ to
\begin{equation}\label{rho}
    \rho = \beta \sinh ^{-1} z
\end{equation}
so that after some straightforward algebra $\widetilde{V} (\rho)$
in eq. (\ref{v-tilde}) reduces to
\begin{equation}\label{v-sech}
    \widetilde{V} (\rho) = \displaystyle \frac{1}{4 \beta ^2} - \left( \mu ^2 -
    \frac{1}{4 \beta ^2} \right) {\rm{sech}} ^2  \frac{\rho}{\beta}
\end{equation}
Thus equation (\ref{schro-const-m}) can be written as
\begin{equation}\label{schro-rho}
    \displaystyle \frac{d^2 \phi}{d \rho ^2} + \left \{ \kappa ^2 +
    \lambda ( \lambda - 1 ) {\rm{sech}} ^2
    \frac{\rho}{\beta} \right\} \phi = 0
\end{equation}
\begin{equation}\label{k}
    {\rm{where}} \qquad \displaystyle \kappa ^2 =  E - \frac{1}{4 \beta ^2}
\end{equation}
and the parameter $\lambda$ depends on the constants $\mu$ and
$\beta$, through the equation
\begin{equation}\label{lambda}
    \lambda ( \lambda - 1 ) = \displaystyle  \mu ^2 -
    \frac{1}{4 \beta ^2}
\end{equation}
For existence of bound states $ \lambda
> 1 $; hence, $ \mid \mu \mid \ >
\displaystyle \frac{1}{2 \beta} $. This gives the permissible
values of $ \lambda $ as
\begin{equation}\label{lambda}
    \lambda = \displaystyle \frac{1}{2} + \frac{1}{2}
    \sqrt{ \displaystyle 1 + 4 \mu ^2 - \frac{1}{\beta ^2}}
\end{equation}

{\begin{figure}[hp]
\begin{center}
\scalebox{0.5}{\includegraphics{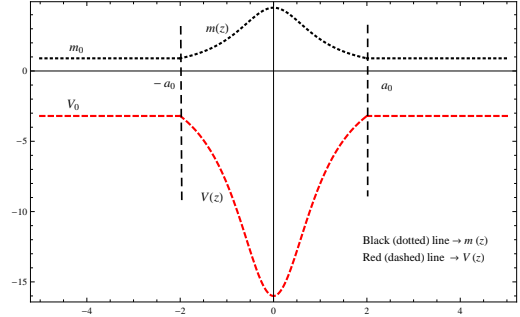}}
\label*{}\caption{\small {Colour online : Plot showing $m(z)$ and
$V(z)$ w.r.t. $z$ }}
\end{center}
\end{figure}}

\noindent For a better understanding of the mass dependence and
the potential in the semiconductor device, we plot $m(z)$ and
$V(z) $ as a function of $z$ in Fig. 1, for a suitable set of
parameter values, viz., $\beta = 4 , \ \mu = 3, \ a_0 = 2 $.


\noindent Let us introduce a new variable
\begin{equation}\label{y}
    y = \displaystyle  \cosh ^2 \frac{\rho}{\beta}
\end{equation}
and write the solutions of (\ref{schro-rho}) as
\begin{equation}\label{phi-u}
    \phi = \displaystyle y^{\frac{ \lambda}{2}} \ u(y)
\end{equation}
In terms of the new variable $y$, equation (\ref{schro-rho})
reduces to the hypergeometric equation
\begin{equation}\label{y-u}
    \begin{array}{lll}
    & & \displaystyle y(1-y) \frac{d^2 u}{dy^2} + \left\{ \left(
    \lambda  + \displaystyle \frac{1}{2} \right)
    -  \left( \lambda  + 1
    \right) y \right\} \displaystyle \frac{du}{dy} \\
    & & - \ \displaystyle \frac{1}{4} \left\{
    \lambda ^2 + \kappa ^2 \beta^2 \right\} u = 0
    \end{array}
\end{equation}
with complete solution \cite{flugge}
\begin{equation}\label{u-hypergeometric}
    \begin{array}{lll}
    u &=& \displaystyle P \ _2F_1 \left( a,b,\frac{1}{2}; 1-y
    \right) + \displaystyle Q (1-y)^{1/2} \cdot \\
    & & \displaystyle _2F_1 \left( a+\frac{1}{2},b + \frac{1}{2},\frac{3}{2}; 1-y
    \right)
    \end{array}
\end{equation}
where $P$ and $Q$ are constants, and the parameters $a$ and $b$
are as defined below :
\begin{equation}\label{ab}
    a = \displaystyle \frac{1}{2} \left( \lambda + i \kappa \beta \right)
    \ \ , \ \ b = \displaystyle \frac{1}{2}
    \left( \lambda - i \kappa \beta \right)
\end{equation}
Thus for the even solution $Q = 0$, whereas for the odd solution
$P = 0$. After some straightforward algebra, the final solution to
the position-dependent mass Schr\"{o}dinger equation (\ref{H-em}),
within the potential well $ -a_0 < z < a_0$, is obtained as
\begin{equation}\label{psi-in}
    \begin{array}{lll}
    \psi _{in} (z) &=& \displaystyle \left( \frac{\beta ^2}{2 (1+z^2)} \right)
    ^{1/4} \left( 1+z^2 \right) ^{\lambda /2} \cdot \\ \\
    & & \displaystyle \left\{
     P \ _2F_1 \left( a,b,\frac{1}{2}; -z^2
    \right) \right. \\
    & & \left. + \  i Q z \
    _2F_1 \displaystyle \left( a + \frac{1}{2},b + \frac{1}{2},\frac{3}{2}; -z^2
    \right) \right\}
    \end{array}
\end{equation}
whereas outside the well ($z  <  -a_0 \ , \ z > a_0 $), the
solutions are given by eq (\ref{psi-out}), with $ k_1 = k_2$.

{\begin{figure}[hp]
\begin{center}
\scalebox{0.4}{\includegraphics{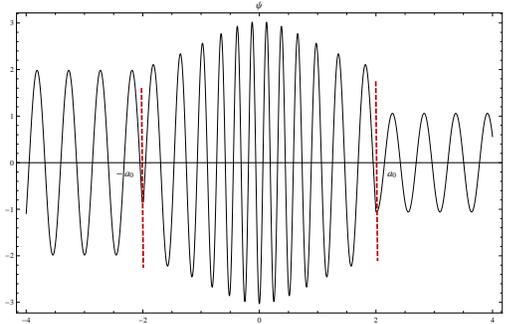}}
\label*{}\caption{\small {Colour online : A plot of Re $ \psi (z)$
vs $z$; Dashed (red) lines show the abrupt heterojunctions }}
\end{center}
\end{figure}}

\noindent For scattering states, $ \kappa ^2$ should be positive,
implying $ E > \displaystyle \frac{1}{4 \beta ^2} $. Now, because
of the spatial dependence of the mass function, the boundary
conditions need to be modified ---
\\ the functions $\displaystyle \psi (z) $ and $\displaystyle
\frac{1}{m(z)} \frac{d \psi (z)}{dz} $ should be continuous at
each heterojunction $\pm a_0$ \cite{benDaniel-Duke,boundary}.
These conditions enable us to evaluate the reflection and
transmission amplitudes $R$ and $T$ respectively.


\noindent The complete solutions for the scattering states in the
entire region $ - \infty < z < \infty $ are plotted in Fig. 2, for
the same set of parameter values as in Fig. 1, viz., $\beta =4 \ ,
\ \mu = 3 \ , \ a_0 = 2 \ , \ E=40 $. The solutions show a
definite nonlinear character inside the well ($-a_0  <  z  <  a_0
$), where the particle mass $m$ is a function of its position $z$.
Thus the effect of the position-dependent mass potential well (in
this particular model) is to introduce a non linear component in
the otherwise plane wave solutions.


\subsection{Case 2 : Potential barrier with position dependent mass}

\noindent As the second example, we consider the region within the
double heterojunction ($a_1 < z < a_2$) to be represented by an
{\it inverted} Morse potential (barrier)
\cite{morse-barrier,zafar-PLA,zafar-PRA} :
\begin{equation}\label{pot-morse}
    V(z) = \left\{
    \begin{array}{lcl}
        & & \displaystyle V_0 \ e^{\alpha z} \left( 2 - e^{
        \alpha z } \right) \qquad  \qquad , \ a_1 < z < a_2 \\
        & & \displaystyle V_0 \ e^{ \alpha a_1} \left( 2 - e^{
        \alpha a_1 } \right) = V_{01} \ , \ - \infty < z < a_1 \\
        & & \displaystyle V_0 \ e^{ \alpha a_2} \left( 2 - e^{
        \alpha a_2 } \right) = V_{02} \ , \ a_2 < z < \infty
    \end{array}
        \right.
\end{equation}
with positive $V_0$. The transmission coefficient of a potential
barrier has wide applications in nuclear fission, heavy-ion
fusion, tunnelling in solids \cite{zafar-PRA}, etc; hence we shall
calculate both the transmission as well as reflection coefficients
and plot them as a function of the energy of the particle. The
Morse barrier potential is particularly useful in investigating
the anharmonicities of the vibrational spectra in molecular and
nuclear physics \cite{zafar-PLA}. For the mass function of the
particle, we consider
\begin{equation}\label{mass-morse}
    m(z) = \left\{
    \begin{array}{lcl}
        & & \displaystyle m_0 \alpha ^2 e^{- 2 \alpha z} \qquad \qquad , \ a_1 < z < a_2 \\
        & & \displaystyle m_0 \alpha ^2 e^{- 2 \alpha a_1} = m_{01} \ , \ - \infty < z < a_1 \\
        & & \displaystyle m_0 \alpha ^2 e^{- 2 \alpha a_2} = m_{02} \ , \ a_2 < z < \infty
    \end{array}
        \right.
\end{equation}

{\begin{figure}[hp]
\begin{center}
\scalebox{0.4}{\includegraphics{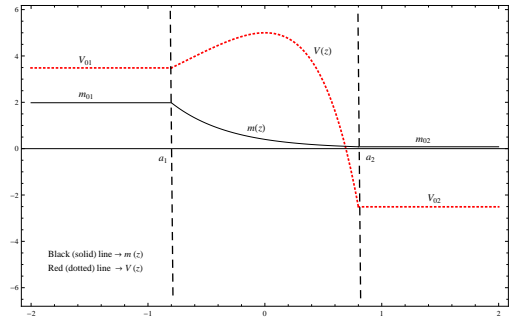}}
\label*{}\caption{\small {Colour online : Plot showing $m(z)$ and
$V(z)$ w.r.t. $z$ , The dashed lines show the abrupt
heterojunctions}}
\end{center}
\end{figure}}


\noindent In Fig. 3, we show the plot of $m(z)$ and $V(z) $ for
this particular model, for a suitable set of parameter values,
viz., $m_0 = 0.4 , \ V_0 = 5 , \ E = 33 $, with the
heterojunctions at $a_1 = -0.8 \ , \ a_2 = 0.8 $.

\noindent For the spatial mass dependence given by eq.
(\ref{mass-morse}), eq. (\ref{psi-z}) transforms the coordinate
$z$ to
\begin{equation}\label{rho}
    \rho = \displaystyle - \sqrt{2 m_0} e^{- \alpha z}
\end{equation}
so that after some straightforward algebra $\widetilde{V} (\rho)$
in eq. (\ref{v-tilde}) reduces to the simple form
\begin{equation}\label{v-tilde-morse}
    \tilde{V} _{\rho} = \displaystyle -
    \frac{2 V_0 \sqrt{2 m_0} }{\rho} - \frac{ 2 m_0 V_0 -
    3/4}{\rho ^2}
\end{equation}
Thus the Schr\"{o}dinger equation for constant mass
(\ref{schro-const-m}) takes the final form
\begin{equation}\label{schro-morse}
    \displaystyle \frac{d^2 \phi}{d \rho ^2} + \left\{ \kappa ^2 +
    \frac{2 m_0 V_0 - 3/4}{\rho ^2} + \frac{2 V_0 \sqrt{2
    m_0}}{\rho} \right\} \phi = 0
\end{equation}
with $\kappa ^2 = E$. To solve equation (\ref{schro-morse}) given
above, let us introduce a new variable
\begin{equation}\label{y-morse}
    y = \displaystyle - 2 i \kappa \rho = \displaystyle i \kappa
     2 \sqrt{2 m_0} e^{- \alpha z}
\end{equation}
in terms of which equation (\ref{schro-morse}) gets simplified to
the form of a Whittaker differential equation \cite{handbook}
\begin{equation}\label{whittaker-eq}
    \displaystyle \frac{d^2 \phi}{d y^2} + \left\{ - \frac{1}{4} +
    \frac{\lambda _1 ^2 + 1/4}{y^2} + \frac{i \lambda _2}{y} \right\} \phi = 0
\end{equation}
with \begin{equation}\label{lambda-12}
    \lambda _1 ^2 = \displaystyle 2 m_0 V_0 - 1 \qquad ,
    \qquad \lambda _2 = \displaystyle \frac{V_0 \sqrt{2 m_0}}{\kappa}
\end{equation}
The solutions of eq (\ref{whittaker-eq}) are given as
\cite{handbook}
\begin{equation}\label{morse-sol}
    \phi = \displaystyle e^{ \pm y/2} y^{ \pm i \lambda _1} M
    \left( a^{\pm}, b^{\pm};y \right)
\end{equation}
where $\displaystyle M
    \left( a^{\pm}, b^{\pm};y \right) $ are the Whittaker
    functions and
\begin{equation}\label{apm-bpm}
    \displaystyle a^{\pm} = \frac{1}{2} \pm i \lambda _1 - i
    \lambda _2 \qquad , \qquad b^{\pm} = 1 \pm 2 i \lambda _1
\end{equation}


\noindent Thus for the entire semiconductor device, the complete
solution for the scattering states in the different regions are
given by
\begin{equation}\label{morse-sol}
\begin{array}{lcl}
    \psi _L  &=& \displaystyle e^{i k_1 z} + R e^{-ik_1 z} \ \ , \  - \infty <
    z < a_1 \\ \\
    \psi_{in} &=& \displaystyle \left( 2 m_0 \right) ^{1/4}
    \sqrt{\alpha}  e^{\frac{- \alpha z }{2}} \left\{ P_1  e^{\frac{y}{2}}
    y ^{i \lambda _1}  M \left(
    a^+ , b^+ ;y \right) \right.\\
    & & \displaystyle  \left. + \ P_2  e^{- \frac{y}{2}}
    y ^{- i \lambda _1}  M \left(
    a^- , b^- ; y \right) \right\} \ , \ a_1 < z < a_2 \\ \\
    \psi _R  &=& T e^{i k_2 z} \ \ \ \  , \  a_2 <
    z < \infty \\
\end{array}
\end{equation}
where the constants $P_1, P_2$ and the reflection and transmission
amplitudes $R$ and $T$ respectively, are determined by matching
the boundary conditions (for pdem systems) at the heterojunctions,
and $y$ and $\rho$ are as defined above.


\noindent The complete solutions for the entire region $ - \infty
< z < \infty $ are plotted in Fig. 4, for a suitable set of
parameter values, viz., $m_0 = 0.4 , \ V_0 = 5 , \ E = 33 , \ a_1
= - 1.5  , \ a_2 = 1.5 $. The solutions show a definite nonlinear
character inside the barrier ($a_1 < z < a_2 $), where the
particle mass is dependent on its position. Thus the effect of the
position-dependent mass barrier (similar to that of the position
dependent potential well in the previous example) is to introduce
a non linear component in the plane wave solutions.


{\begin{figure}[hp]
\begin{center}
\scalebox{0.55}{\includegraphics{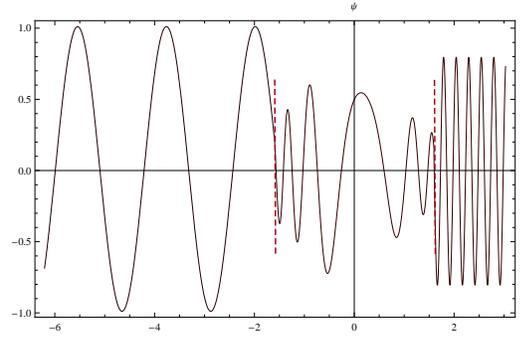}}
\label*{}\caption{\small {Colour online : A plot of Re $ \psi (z)$
vs $z$; Dashed (red) lines show the abrupt heterojunctions }}
\end{center}
\end{figure}}

\section{Conclusions and Discussions}

\noindent To conclude, we obtained the exact analytical solutions
for the scattering states of a particle (electron or hole) inside
a semiconductor device with a double heterojunction, when the mass
of the particle is assumed to be dependent on its position inside
the heterojunctions, but constant outside. We studied two explicit
models in this work
--- one a pdem diffused potential well, the other
a pdem potential barrier (Morse barrier). In each case it is
observed that the effect of the spatial dependence on the particle
mass is to introduce a non linear component in the otherwise plane
wave solutions.

{\begin{figure}[hp]
\begin{center}
\scalebox{0.8}{\includegraphics{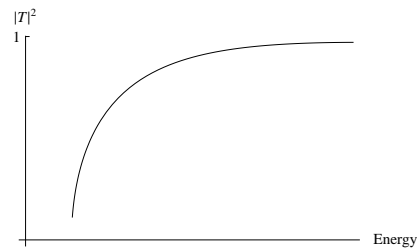}}
\label*{}\caption{\small {Plot of $|T|^2$ vs $E$ }}
\end{center}
\end{figure}}

\pagebreak

{\begin{figure}[hp]
\begin{center}
\scalebox{0.8}{\includegraphics{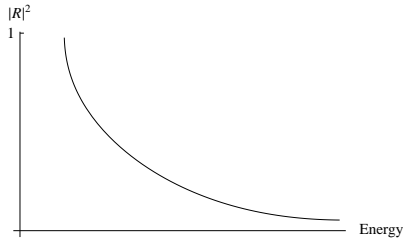}}
\label*{}\caption{\small {Plot of $|R|^2$ vs $E$ }}
\end{center}
\end{figure}}

We also calculated the transmission and reflection coefficients,
$|T|^2$ and $|R|^2$ respectively, for the potentials studied here.
These are plotted in Fig. 5 and Fig. 6 respectively, as a function
of the energy $E$ for the pdem particle in a Morse barrier. As the
energy increases, the transmission coefficient also increases,
finally reaching unity, whereas the reflection coefficient follows
the reverse trend and goes to zero. This observation is similar to
that in ref. \cite{bianchi}, where the authors show that the
transmission coefficient for the one-dimensional scattering
problem with pdem normally tends to unity as energy goes to
infinity, provided the mass is a continuous function of position.


This simple, yet straightforward approach is just a way of
understanding basic physics of the electronic properties of a
semiconductor device, comprising of a double heterojunction, where
the intermediate layer is sufficiently thin for quantum properties
to be exhibited. It is expected that the observations made in this
work will provide some useful insight in studies related to
electron transport in semiconductor heterostructures, i.e. in the
physical properties of such materials. Actual materials are made
up of a large number of atomic potentials. Nevertheless, the
crystal potential may be approximated by a single potential ---
the global average of the individual potentials, and this approach
would still be valid. However, for extremely thin intermediate
layer, the individual potentials may become significant enough for
this approximation to break down. This calls for a more rigorous
approach, and we propose to take up its study in the near future.



\section{Acknowledgement}

The author thanks P. Roy for some very fruitful discussions.
Financial assistance for the work was provided for by the Dept. of
Science and Technology, Govt. of India, through its grant
SR/WOS-A/PS-06/2008. Thanks are also due to the unknown referee
for some useful comments.


\end{document}